\documentclass[a4paper,fleqn,usenatbib]{mnras}
\usepackage{newtxtext,newtxmath}

\usepackage[T1]{fontenc}
\usepackage{ae,aecompl}

\usepackage{graphicx}	
\usepackage{amsmath}	
\usepackage{amssymb}	

\title[SMFs of different kinematic classes]{The SAMI Galaxy Survey: 
The contribution of different kinematic classes to the stellar mass function of nearby galaxies}

\author[K. Guo et al.]{
Kexin Guo,$^{1,2}$\thanks{E-mail: kxguo@pku.edu.cn}
Luca Cortese,$^{2,3}$
Danail Obreschkow,$^{2,3}$
Barbara Catinella,$^{2,3}$
\newauthor
Jesse van de Sande,$^{3,4}$
Scott M. Croom,$^{3,4}$
Sarah Brough,$^{5}$
Sarah Sweet,$^{3,6}$
\newauthor
Julia J. Bryant,$^{3,4,7}$
Anne Medling,$^{8}$
Joss Bland-Hawthorn,$^{4}$
Matt Owers,$^{9,10}$
\newauthor
Samuel N. Richards$^{11}$
\\
$^{1}$Kavli Institute for Astronomy and Astrophysics, Peking University, 5 Yiheyuan Road, Haidian District, Beijing 100871, P.R.China\\
$^{2}$International Centre for Radio Astronomy Research (ICRAR), University of Western Australia, Crawley, WA 6009, Australia\\
$^{3}$ARC Centre of Excellence for All Sky Astrophysics in 3 Dimensions (ASTRO 3D),Australia\\
$^{4}$Sydney Institute for Astronomy, School of Physics (SIfA), A28, The University of Sydney, NSW, 2006, Australia\\
$^{5}$School of Physics, University of New South Wales, NSW 2052, Australia\\
$^{6}$Centre for Astrophysics and Supercomputing, Swinburne University of Technology, PO Box 218, Hawthorn, VIC 3122, Australia\\
$^{7}$Australian Astronomical Optics, AAO-USydney, School of Physics, University of Sydney, NSW 2006, Australia\\
$^{8}$Ritter Astrophysical Research Center, University of Toledo Toledo, OH 43606, USA\\
$^{9}$Department of Physics and Astronomy, Macquarie University, NSW 2109, Australia\\
$^{10}$Astronomy, Astrophysics and Astrophotonics Research Centre, Macquarie University, Sydney, NSW 2109, Australia\\
$^{11}$SOFIA Science Center, USRA, NASA Ames Research Center, Building N232, M/S 232-12, P.O. Box 1, Moffett Field, CA 94035-0001, USA\\
}

\date{Accepted XXX. Received YYY; in original form ZZZ}

\pubyear{2018}
\begin{document}
\label{firstpage}
\pagerange{\pageref{firstpage}--\pageref{lastpage}}
\maketitle

\begin{abstract}
We use the complete Sydney-AAO Multi-object Integral field spectrograph (SAMI) Galaxy Survey
to determine the contribution of slow rotators, as well as different types of fast rotators,
to the stellar mass function of galaxies in the local Universe.
We use stellar kinematics not only to discriminate between fast and slow rotators, 
but also to distinguish between dynamically cold systems (i.e., consistent with intrinsic axis ratios$<0.3$) and systems including a prominent dispersion-supported bulge.
We show that fast rotators account for more than $80\%$ of the stellar mass budget of nearby galaxies,
confirming that their number density overwhelms that of slow rotators at almost all masses from $10^{9}$ to $10^{11.5}{\rm M_\odot}$.
Most importantly, dynamically cold disks contribute to at least $25\%$ of the stellar mass budget of the local Universe, significantly higher than what is estimated from visual morphology alone.
For stellar masses up to $10^{10.5}{\rm M_\odot}$, this class makes up $>=30\%$ of the galaxy population in each stellar mass bin. The fact that many galaxies that are visually classified as having two-components have stellar spin consistent with dynamically cold disks suggests that
the inner component is either rotationally-dominated (e.g., bar, pseudo-bulge) 
or has little effect on the global stellar kinematics of galaxies.
\end{abstract}


\clearpage



\section{Introduction}
The stellar mass function (SMF) of galaxies - i.e., the number density of galaxies per unit of stellar mass - 
has become a key tool for galaxy evolution studies,
as its shape and normalization are regulated by the mass assembly history of galaxies \citep[e.g.,][]{Mac10,Men12,Kang13}.
With increasing observational data on large samples of galaxies in both the local and the high-redshift Universe,
considerable efforts have been made to quantify the overall shape of the SMF
for different populations of galaxies as well as its evolution over cosmic time \citep[e.g.,][]{Ilbert10,Peng10,Baldry12,Davidzon17}.
This has also made the SMF the primary tool for calibrating cosmological simulations of galaxy formation and evolution 
\citep[e.g.][]{Genel14,Crain15,Pill18}.

%
%
The analysis of the SMF is even more powerful when applied to different galaxy classes,
as it encapsulates the effects of a variety of physical processes on the mass accretion history of galaxies. Thus, it provides important clues on the mass regimes within which different evolutionary paths for galaxy transformation are most likely.
For instance, 
the comparison between SMFs for star-forming galaxies (SFG) and passive galaxies suggests a faster quenching rate in more massive galaxies \citep[e.g.,][]{Peng10,Ilbert10}.

Among these studies, the study of SMFs of galaxies of different morphology is a topic of particular interest, as morphological transformation is expected to be intimately tied to galaxy evolution and accretion history. Disc-like structures mostly arise from dissipational gas accretion \citep[e.g.,][]{FE80}, while the formation of spheroidal structures, including galaxy bulges and ellipticals, has been widely believed to be related to dissipationless mergers \citep[e.g.,][]{Cole00}. Disk mergers also have an impact in torquing gas to lose the angular momentum in transforming a galaxy from late-type to early-type \citep[e.g.,][]{Ger81,Hern92,Hern93,Heyl94}.

Several independent works have shown that the mass distribution of late-type (disk-dominated) and early-type (bulge-dominated) galaxies show distinct differences.
The early-types dominate at $M_*>10^{10}{\rm M_\odot}$, while the late-types dominate at lower stellar mass \citep{Kel14,Mof16a}. Despite this, the two classes contribute nearly equally to the global stellar mass budget of the local Universe, $\Omega_{*}$ \citep{Driver07, Mof16b}. All these results could implicate that mergers play as important a role in mass assembly as gas accretion, and dominate the galaxy evolution in the massive regime.


The underlying assumption of the above scenario is that visual morphology is able to isolate truly different kinematic components, especially components like dispersion-supported bulges and rotationally-supported disks, which are directly linked to the merging history of galaxies. 


While this assumption was commonly accepted in the past, the last decade has clearly shown that galaxies with distinct visual morphology can share similar kinematic properties.
Most early-type galaxies (ETGs) are found to have a rotational component \citep[e.g.,][]{KB96,Faber97,Cap07,Ems07,Ems11}, arguing against the origin of ETGs or massive galaxies as exclusively (dissipationless) mergers. Non-rotating galaxies, on the other hand, make up only $\leq1/3$ of the total ETGs, and are found almost only at stellar mass $M_*>10^{11}{\rm M_\odot}$ \citep{Cap16,Brough17,Graham18}.

Therefore, the contribution of the different physical processes to the mass assembly history of galaxies is no longer well understood.
While major mergers or multiple minor dry mergers are necessary for the formation of slow rotators (SRs) \citep[e.g.,][]{Naab14},
fast rotators (FRs) can be transformed from late-type disks following a continuous kinematic evolution \citep{Cortese16, Cortese19}.
Additionally, the bulge component in late-type galaxies can either be a dispersion-dominated classical bulge arisen from galaxy mergers,
or a rotating system originated from the secular evolution of disks \citep{Men14, Erwin15}.

The advent of integral-field spectroscopic (IFS) galaxy surveys is finally helping us to shed some light on this issue, providing a more physically-motivated approach to investigate the assembly history of galaxies. 
Many works have focused on the relative fraction between FRs and SRs \citep[][and references therein]{Cap16}, and in most cases limited to pre-visually-classified early-type galaxies. 

In this study, we take a different approach and present the first investigation of the contribution of different kinematic families of galaxies to the stellar mass function in the local Universe. In addition to presenting the first stellar mass functions for slow and fast rotators, we split FRs into `dynamically cold' FRs (i.e., consistent with intrinsic axis ratios smaller than $\sim$0.3) and `composite' rotators (i.e., either thick disks or bulge-plus-disk objects) to determine their overall contribution to the stellar mass function. This allows us to provide new constraints on the mass-dependent contribution of different physical processes to the whole population of galaxies.

This paper is structured as follows. We introduce the dataset in Section 2
and describe the SMF fitting method in Section 3. 
We show our results in Section 4 and discuss in Section 5.
A Chabrier initial mass function \citep[IMF;][]{Chabrier03} is used throughout the paper.
We assume the following cosmological parameters: $\Omega_0=0.3, \Omega_\Lambda=0.7, H_0=70\,\rm {km\,s^{-1}}$.

\begin{figure}
 \includegraphics[width=\columnwidth]{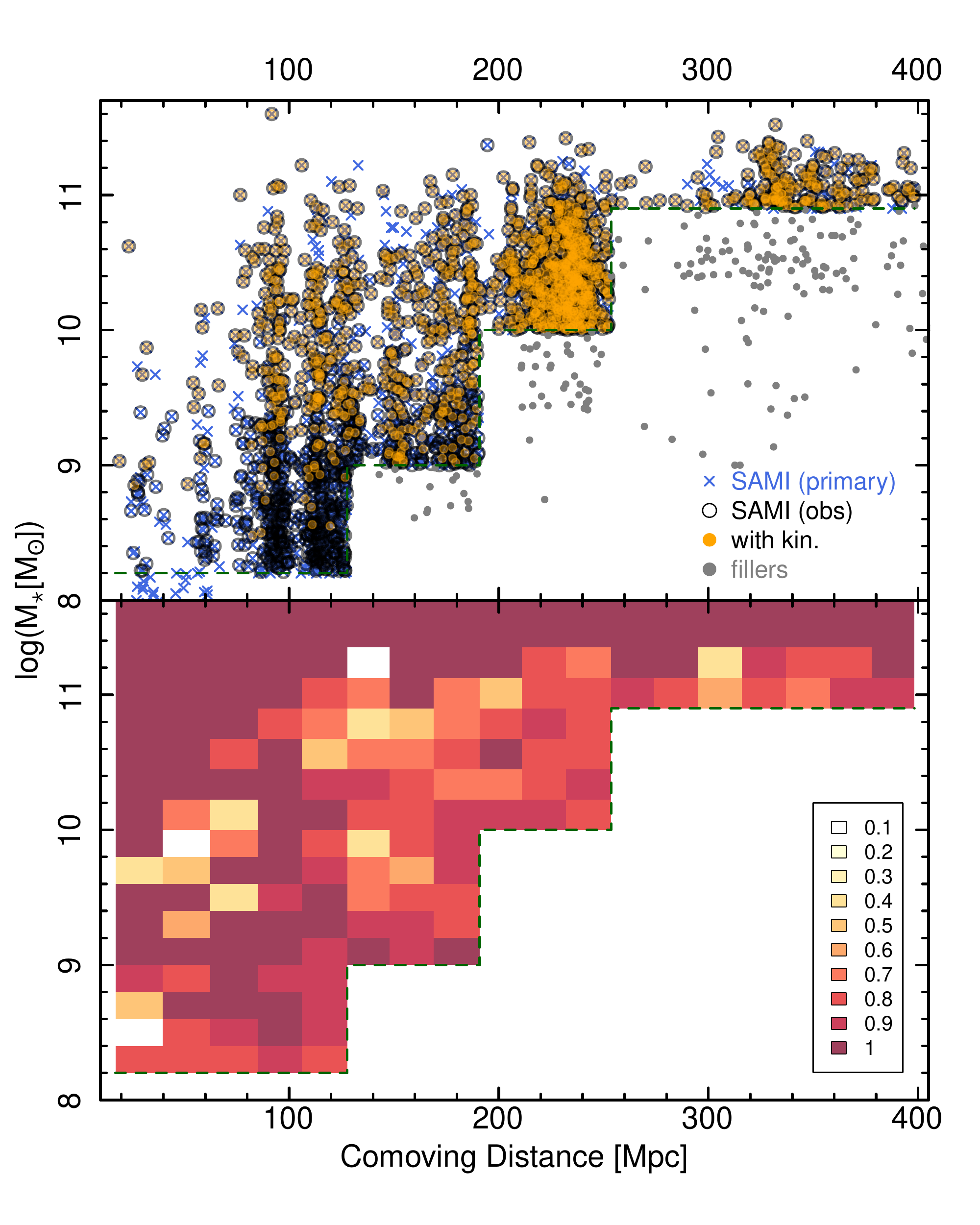}
 \caption{
 The distribution ({\it upper}) and the completeness ({\it lower}) of SAMI(obs) in the stellar mass-comoving distance plane.
 In the upper panel, the distribution of the SAMI primary sample are labeld as {\it blue crosses}. All galaxies observed by SAMI are labeld as {\it black empty circles} (SAMI(obs)), except the filler galaxies ({\it grey points}). Galaxies with available kinematics in SAMI(obs) are shown as {\it orange points}. In the lower panel, 1 is assigned to grids with no galaxy in either SAMI(PS) or SAMI(obs) since the fraction-weighted incompleteness correction is not applicable (\S3.1).}
 \label{fig1}
\end{figure}

\section{Data}
Our sample is extracted from the overlap between the SAMI Galaxy Survey internal full data release (v0.11) and the Galaxy and Mass Assembly (GAMA) survey \citep{Driver11}. 
The SAMI Galaxy Survey targeted 2153 low-redshift galaxies ($z<0.1$) in the 3 GAMA equatorial regions corresponding to an area of $144\,{\rm deg}^2$ in total,
using the SAMI multi-object IFS instrument \citep{Croom12} mounted on the 3.9 metre Anglo-Australian Telescope \citep{Bryant15}.
SAMI fibres are fed to the double-beam AAOmega spectrograph \citep{Sharp06}, 
providing spectral resolutions of $R=1812$ for the blue part ($3700-5700$\AA),
and $R=4263$ for the red part ($6300-7400$\AA) of the spectrum, respectively.
The SAMI cubes have 50$\times$50 0.25 $(0.5\times0.5)\,{\rm arcsec}^2$ spaxels along 
the spatial direction, covering the 14.7 arcsec diameter aperture of the SAMI hexabundle \citep{BH11,Bryant14} and an average seeing of 2.16 arcsec.
The data reduction is described in detail in \citet{Allen15} and \citet{Sharp15}.
We refer readers to \citet{Green18} and \citet{Scott18} for Data Releases I and II, respectively.

In this work, we focus on the 1896 galaxies included in the
SAMI primary sample (i.e., excluding filler targets), in the footprint of GAMA as described in \citet{Bryant15}, and with a lower-limit mass cut at $M_*>10^{8.2}{\rm M_\odot}$ (hereafter ``SAMI(obs)'', {\it black empty circles} in Figure~\ref{fig1})
\footnote{We decide to remove 45 galaxies with stellar mass ranging from $10^{7.42}$ to $10^{8.2}{\rm M_\odot}$  from the original catalog for two reasons. One is that the statistics is low in this relatively large range ($\sim$0.8\,dex). The second is that the SAMI sample selection for $M_*<10^{8.2}{\rm M_\odot}$ is not stepwise and the incompleteness is hard to constrain.}.
As described in \citet{Bryant15}, the configuration of SAMI plates is done to maximize the number of objects observable within a SAMI field of view, without pre-selection on morphology or environment. The stellar mass estimation is described in \citet{Bryant15}, following the spectral energy distribution (SED) fitting method of \citet{Taylor11}. The uncertainty introduced by photometric error is around $0.05\,{\rm dex}$, while the intrinsic scatter of colour-dependent M/L is about $0.1\,{\rm dex}$.





Visual morphology classification in SAMI has been performed 
taking advantage of Sloan Digital Sky Survey Data Release Nine \citep[SDSS DR9;][]{Ahn12} {\it g\,r\,i} colour images, as discussed in \citet{Cortese16}.
Except for 80 unclassified galaxies (``?" in Figure~\ref{fig4}) and galaxies without a consensus in morphology classification (``NA''), 
1816 galaxies have classifications from late-type disks to ellipticals without ambiguity.
Briefly, galaxies with early-type morphologies (round and smooth) or no presence of spiral arms are classified as ETGs, and the galaxies fulfilling these conditions but having signs of star formation are excluded. ETGs with disks are further classified as S0s.
Here, we constrain LTGs to be only visually pure disks/irregular galaxies (late-type spirals, ``LS") and classify other late-type galaxies with intermediate types of morphologies as MTGs.
Precisely, LTGs ({\it blue}) corresponds to ``LS'' in Figure~\ref{fig4},
and MTGs ({\it greens}) consists of early-or-late-type spirals (``ES/LS'') and early-type spirals (``ES''), with ETGs ({\it red} to {\it yellow}) including all earlier types.

Effective radii ($r_e$), ellipticities ($\epsilon$), and position angles have been derived 
using the Multi-Gaussian Expansion \citep[MGE;][]{EMB94,Cap02} 
technique performed on SDSS $r$-band images with the code from \citet{Scott09} and D'Eugenio et al. (in prep). 
The MGE method has been performed to fit the observed surface brightness profile in terms of the sum of two-dimensional Gaussians while keeping the position angles of the Gaussians constant and taking the point spread function (PSF) into account. $r_e$ is defined as the semi-major axis effective radius measured from the best MGE fit, and the ellipticity of the galaxy within one $r_e$ is the $\epsilon$ used in this study.

Stellar kinematics are measured from the SAMI data by 
using the penalized pixel fitting code (pPXF, \citealt{CE04, Cap17}) as described in \citet{Vds17a}.
1222 Galaxies in the sample have stellar kinematic information (aperture uncorrected) 
following the criteria of \citet{Vds17b}:
signal-to-noise (S/N)>3\AA$^{-1}$, $\sigma_{\rm obs}>$FWHM$_{\rm intr}/2\sim35$km\,s$^{-1}$,
and $\sigma_{\rm error}<\sigma_{\rm obs}*0.1+25$km\,s$^{-1}$ \citep[see][]{Vds17a}
\footnote{Only 28 galaxies have ${\rm r_e<HWHM_{PSF}}$ and removing them does not affect our results.}.
No additional cut on stellar mass or morphology or other kinematic features is applied.
The spin parameter within one ${\rm r_e}$, $\lambda_{\rm r_e}$ for each galaxy has been calculated based on \citet{Ems07,Ems11}, according to Equation 9 in \citet{Cortese16}:
\begin{equation}
 \lambda_{R} \equiv \frac{\langle R|V|\rangle}{\langle R\sqrt{V^2+\sigma^2} \rangle}=
 \frac{\sum_{k=1}^{n}F_kR_k|V_{k\,\rm los}|}{\sum_{k=1}^{n}F_kR_k\sqrt{V_{k\,\rm los}^2+\sigma_k^2}} ,
\end{equation}
where 
$n$ includes all pixels within ${\rm r_e}$.

To summarize, we focus on 1896 galaxies observed and selected (``SAMI(obs)'') in total, of which 1816 have certain morphologies from late to early type, and 1222 have kinematic information.

\begin{figure*}
 \includegraphics[width=\linewidth]{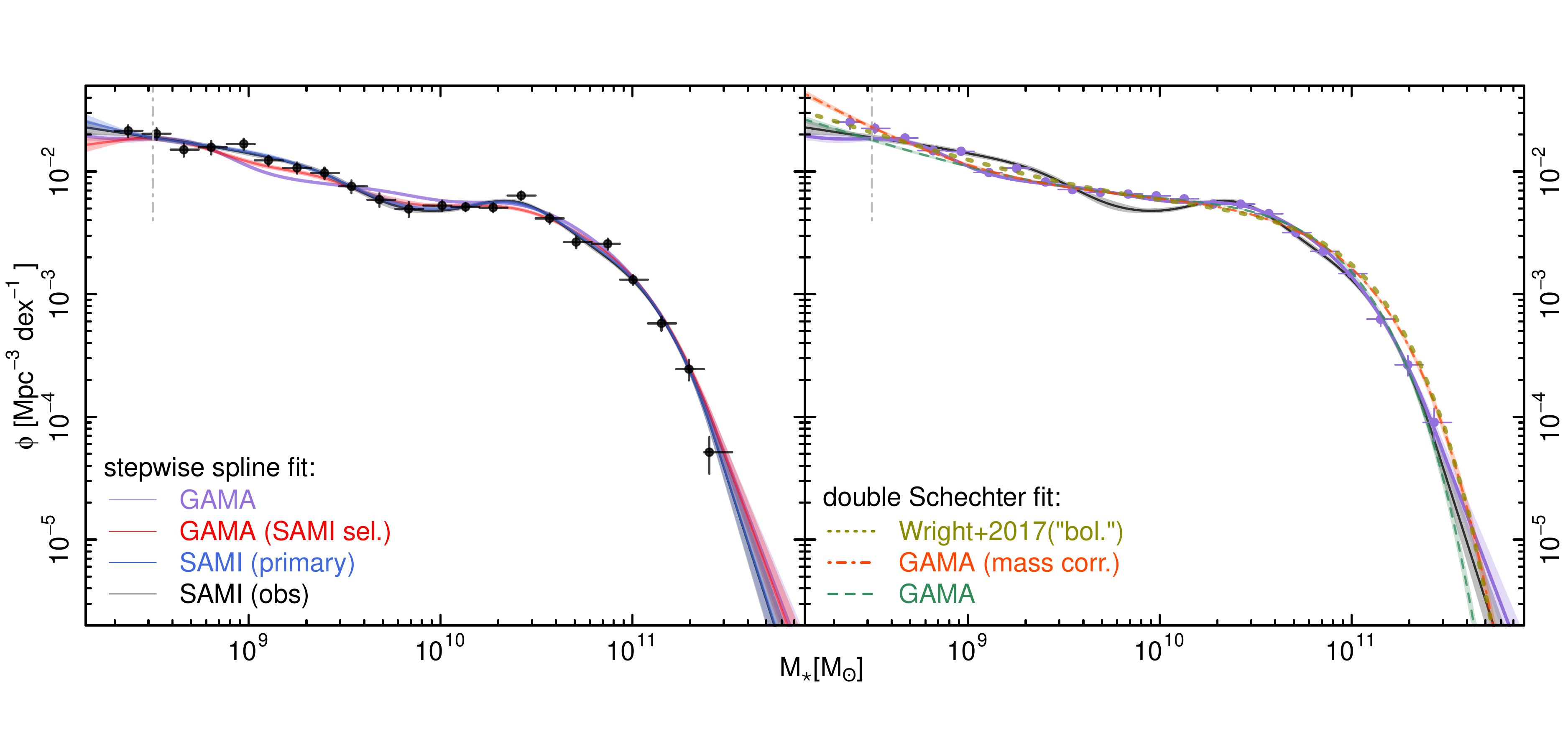}
 \caption{
 The comparison of the SMFs for GAMA and SAMI galaxies obtained using different sample selections and fitting techniques. {\it Left}: The coloured lines show the SMF obtained using a stepwise function for SAMI(obs) ({\it black}), the GAMA-I sample defined in \S3 ({\it purple}), a SAMI-like selection applied to the GAMA-I sample ({\it red}) and the SAMI primary sample. The {\it black points} show the SAMI (obs) SMF per bin of stellar mass. {\it Right}: The SMFs obtained from fitting a double-Schechter functions for the  GAMA-I sample with and without aperture corrections applied to the stellar mass ({\it red} and {\it green}) and the SMF from Wright et al. (2017) based on the GAMA sample with MAGPHYS stellar masses ({\it dotted olive line}). For comparison, the step-wise fits for SAMI(obs) and GAMA-I are overplotted in {\it black} and {\it purple}. The {\it purple points} show the GAMA-I SMF per bin of stellar mass. In both panels, the vertical {\it grey dash-dotted lines} show the lower limit adopted for estimating $\Omega_{*}$ (see text for more details). The consistency between the {\it solid black line} and the SMFs obtained using different sample selections and/or fitting techniques confirms the reliability of our method and incompleteness correction.
 }
 \label{fig2}
\end{figure*}

\section{Method}
To derive the SMFs for our sample we use the fitting tool `dftools' \citep{Ob18}.
`dftools' provides maximum-likelihood fitting with both Schechter \citep{Schechter76} and stepwise (either linear or spline) functions,
with over/under-density estimated and corrected automatically and iteratively.
Similar to other SMF fitting methods,
the package estimates an effective volume for each grid of $M_*$, 
based on a selection function describing the ratio between the number of galaxies accounted into the fitting and that in the real Universe, at given $M_*$ and redshift.
However, since the SAMI galaxies are selected from the GAMA I survey which consists of 3 different fields (G09, G12 and G15), the over/under-density correction does not only depend on redshift
but also deviates from field to field.
Thus, the maximum volume that a galaxy can be observed after density correction (effective volume $V_{\rm eff}$) at given stellar mass is different from field to field.
To address this complication, instead of running the fitting procedure over the full sample,
we calculate the effective volume for individual galaxies and incorporate them directly into the fitting. In this section, we describe our method of correcting for the density inhomogeneity and 
the incompleteness of the sample.
We also compare the SMF derived based on this method with that from a volume-limited sample and that from previous works.

\subsection{A grid-based $V_{\rm eff}$ method}
Although `dftools' fits SMFs to datasets without binning the data, we evaluate the effective volume as a function of stellar mass and cosmic distance on a fine grid (0.2 dex x 19 Mpc cells) for practical convenience.
We adopt the calculation of effective volume in \citet{Baldry12}, 
which is equivalent to the density-corrected $V_{\rm max}'$ method in \citet{Cole11}.
Practically, we select galaxies in the GAMA-I DR2 catalogue \citep{Liske15} with $M_*>10^{9.6}{\rm M_\odot}$
(corresponding to $M_r<-18$\,mag) as the density-defining population (DDP),
and the effective volume for each galaxy $i$ is given by:
\begin{equation}
 V_{\rm eff,i}=\frac{\rho_{\rm ddp}(0.004;z_{\rm max,i})}{\rho_{\rm ddp}(0.004;0.095)},
\end{equation}
where $\rho_{\rm ddp}(z_a;z_b)$ is the number density of DDP between redshift $z_a$ and $z_b$, 
and 0.004 and 0.095 are lower and upper redshift limits for SAMI galaxies.
Given SAMI's stepped sample selection (Figure~\ref{fig1}),
$z_{\rm max,i}$ is one of [0.004, 0.02, 0.03, 0.045, 0.06, 0.095].
It should be noted that,
to recover the under-densities and cosmic variance of low-redshift GAMA regions \citep{Driver11},
we scale our derived SMFs up by a factor of 1.13.
This factor is applied to match
the integrated number density of galaxies with $M_*>10^{10}{\rm M_\odot}$ 
with that calculated based on GAMA galaxies in the same mass range within $0.07<z<0.19$,
the value of which is also adopted as the fiducial density in \citet{Wright17}.

Besides the galaxy distribution in real space,
the incompleteness of the observations is an additional factor that needs to be accounted for.
The simple approach that we apply is to modify the effective volumes in grids of mass and comoving distance
by multiplying the original $V_{\rm eff,i}$ by a completeness factor, 
described by the fraction of galaxies observed in each grid,
i.e., $\frac{N_{\rm obs}}{N_{\rm PS}}$,
where $N_{\rm obs}$ and $N_{\rm PS}$ are the number of galaxies observed and 
that in SAMI primary sample, respectively.
No correction is applied to grids with no galaxy in either sample.
The distribution of the completeness factor is shown together 
with the stepwise-selected SAMI galaxies in Figure~\ref{fig1}.
The resulting SMF fitted based on the SAMI (obs) sample ({\it black}) is shown together 
with that of the SAMI primary sample ({\it blue}) in Figure~\ref{fig2}.
The consistence between them for $M_*>10^{8.5}\rm M_\odot$ suggests that the error in grid-based incompleteness correction (Figure~\ref{fig1}) is negligible.

To assess the reliability of reproducing SMFs based on galaxies 
selected from an original dataset,
we have performed multiple experiments with artificial selection functions,
and compared the output SMFs with the one fit from a parent sample,
which consists of galaxies between $0.004<z<0.095$ 
with $\log(M_*/{\rm M_\odot})>7.6+7.7\times10^{-3}\cdot(r/{\rm Mpc})-6.6\times10^{-6}\cdot(r/{\rm Mpc})^2$ 
in the GAMA-I DR2 catalog \citep{Taylor11}.
The above criterion is defined from a binomial fitting to the peak value in the distribution of $M_*$ 
as a function of comoving distance $r$.
We find that a stepwise-spline function with a `reasonable' step width 
\footnote{For example, the width should not be smaller than 0.1\,dex 
in fitting a SMF based on 300 galaxies ranging from $10^9$ to $10^{12}{\rm M_\odot}$.
In practice, we adopt a step width of 0.34 dex to make sure that every step contains >25 galaxies.}
works better than a (double) Schechter function in reproducing SMFs.
That is because a stepwise-spline function
is a non-parametric model (in the limit of small enough steps) that is more generic than Schechter functions, for which there is no straightforward justification when considering sub-populations of galaxies. 

In Figure~\ref{fig2},
we show the comparison between the SMF fitted by double Schechter function 
based on the GAMA parent sample ({\it right panel, green dashed}),
and the SMF fitted by step-wise functions based on the same sample ({\it purple solid}). 
The difference between the two is <1\% in both number and mass density for galaxies with $M_*>10^{8.2}\,{\rm M_\odot}$.
The SMF fitted by step-wise functions based on GAMA galaxies selected according to the same step-series selection criteria of SAMI primary sample (``GAMA(SAMI sel.)'', {\it red}) is also plotted.
The differences between the GAMA(SAMI sel.) SMF and the former two GAMA SMFs are both <2\% (6\%) in number (mass) density at the same mass range.
This error from the stepwise selection is numerically smaller than the $1\sigma$ uncertainty of fitting 
(note that this is not an indication of overfitting since errors can be correlated between different mass bins).

The stellar mass estimations adopted by the SAMI and GAMA surveys are slightly different,
with the former based on a $g-i$ colour dependent $M_*/M_i$ \citep{Bryant15},
approximating the spectral energy distribution (SED) fitting method of \citet{Taylor11} 
that have been applied to the latter.
The resulting difference in SMF fitting, 
between SAMI primary sample (``SAMI(primary)'', {\it blue}) and GAMA galaxies selected by SAMI selection criteria ({\it red}),
can also be found in the left panel of Figure~\ref{fig2}.
The two SMFs match each other very well (<2\% in both number and mass density) for $M_*>10^{9.5}{\rm M_\odot}$,
and the deviation for the less massive part causes an offset of 8\% in number density but less than 1\% in mass density,
which is very small compared to other uncertainties contributed by selection function and fitting errors.

Correction of Eddington bias \citep{Eddington13} is a built-in procedure in `dftools'.
However, given the low level of random noise ($\sim0.05$dex) in stellar mass estimation 
propagated from the photometric error,
the uncertainty in fitting parameters caused by not considering Eddington bias 
is small relative to that of the fitting itself,
which is estimated by Laplace approximation, 
equivalent to performing bootstrap iterations of resampling in most cases \citep[for details, see][]{Ob18}.

We find a total value of $\Omega_{*}=1.41\pm0.10\substack{+0.36 \\ -0.29}\times 10^{-3}$ relative to the critical density,
calculated by integrating the mass density of galaxies with $10^{8.5}<M_*/{\rm M_\odot}<10^{12}$,
where the second uncertainty components come from the systematic uncertainty of $0.1\,{\rm dex}$ 
on stellar mass estimation based on {\it g-} and {\it i-} band photometry \citep{Taylor11}.
The difference caused by not applying aperture correction for stellar mass estimation 
({\it green dashed} vs. {\it orange dash-dotted} in the right panel of Figure~\ref{fig2}) is $\sim 4.4\%$ in this mass range.
Given the value of $\Omega_{*}=1.61\times10^{-3}$ by integrating 
the ``bolometric'' SMF of \citet{Wright17} in the same mass range,
the difference between our value and that from other works \citep[][and references therein]{Wright17}
can mostly be explained by the difference in sample selection and fitting methods.

To summarize, despite the marginal difference caused by using different stellar mass catalogues, the reliability of our method in correcting the selection function and other incompleteness has been demonstrated by the consistency between the SMF estimated based on the GAMA-I sample and that of galaxies observed by SAMI. Given that our main goal is to compare SMFs and contributions to $\Omega_{*}$ for different types of galaxies (i.e., relative differences), small systematic effects in the normalisation do not significantly affect our main conclusions.

\subsection{An application on stepwise SMFs divided by morphology}
We apply the grid-based $V_{\rm eff}$ method described above in deriving SMFs 
divided by morphology.
Morphological types are defined in the previous section, and include ETGs (ellipticals and S0s), 
MTGs (mainly early spirals), and LTGs (late spirals).
The fraction of galaxies that are morphologically classified is higher than 97\%. 
Practically, we correct this incompleteness by weighting $V_{\rm eff}$ by the fraction of galaxies with certain morphologies in grids defined in \S3.1. 
Only marginal differences would be found if no correction was applied.

\begin{figure}
 \includegraphics[width=\columnwidth]{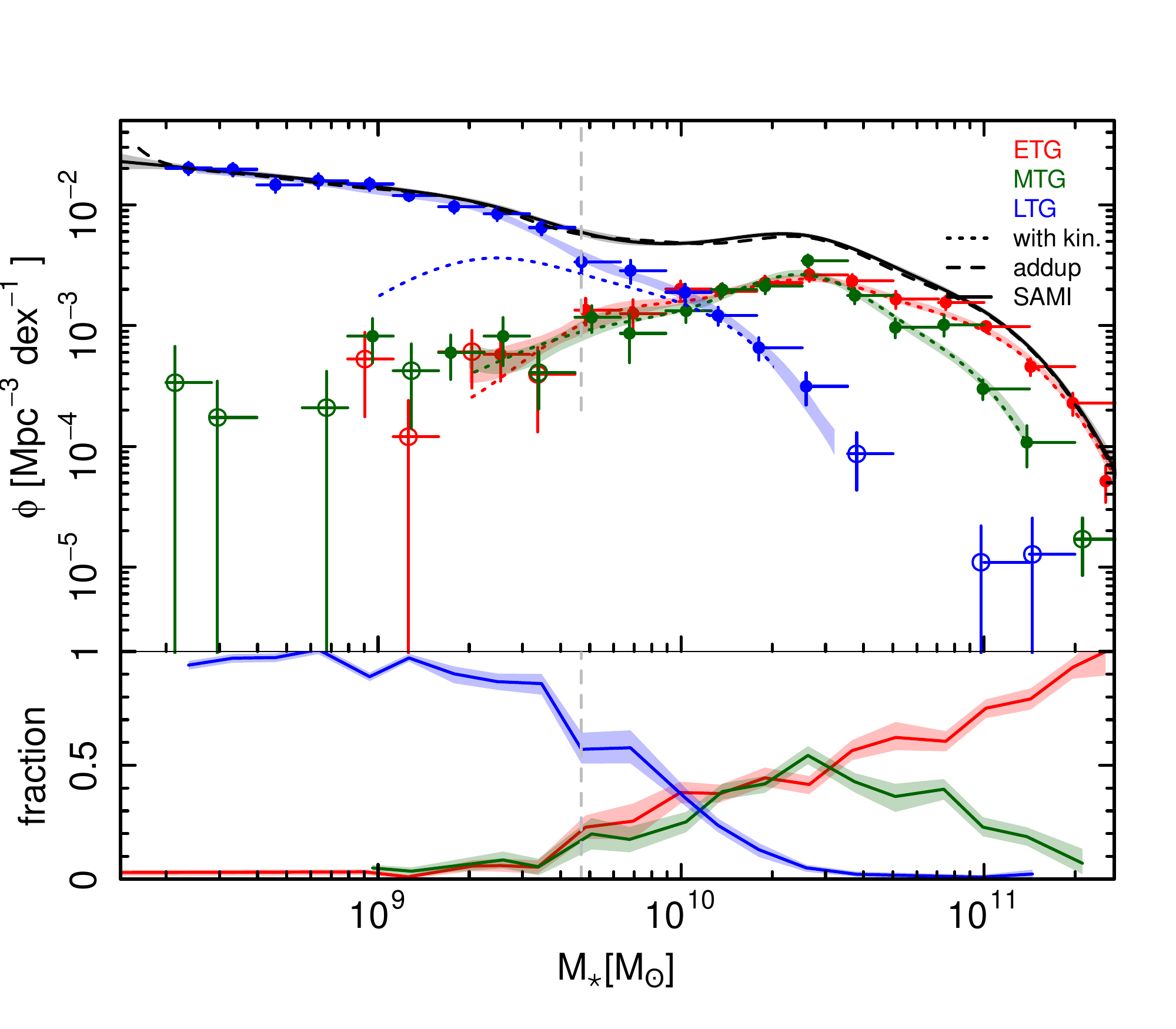}
 \caption{{\it Upper panel:} The SMFs for all our SAMI observed sample ({\it black solid line}), and for individual classes of visual morphology (from ETG in red to LTG in blue).
 We show the best-fitting function for bins with more than 4 galaxies.
 The {\it dotted} lines show how the result would change if we included only galaxies with available kinematic information.
 The sum of the SMFs of all fitting curves is shown by the {\it black dashed line}.
 {\it Lower panel:}  The fraction in the number density of different classes at given stellar mass.
 The {\it vertical dashed line} shows the completeness limit in stellar mass when comparing galaxies of different kinematic type, same as that in Figure~\ref{fig5}b.
 Our SMFs for different morphological classes are qualitatively consistent with previous works.}
 \label{fig3}
\end{figure}

\begin{figure}
 \includegraphics[width=\columnwidth]{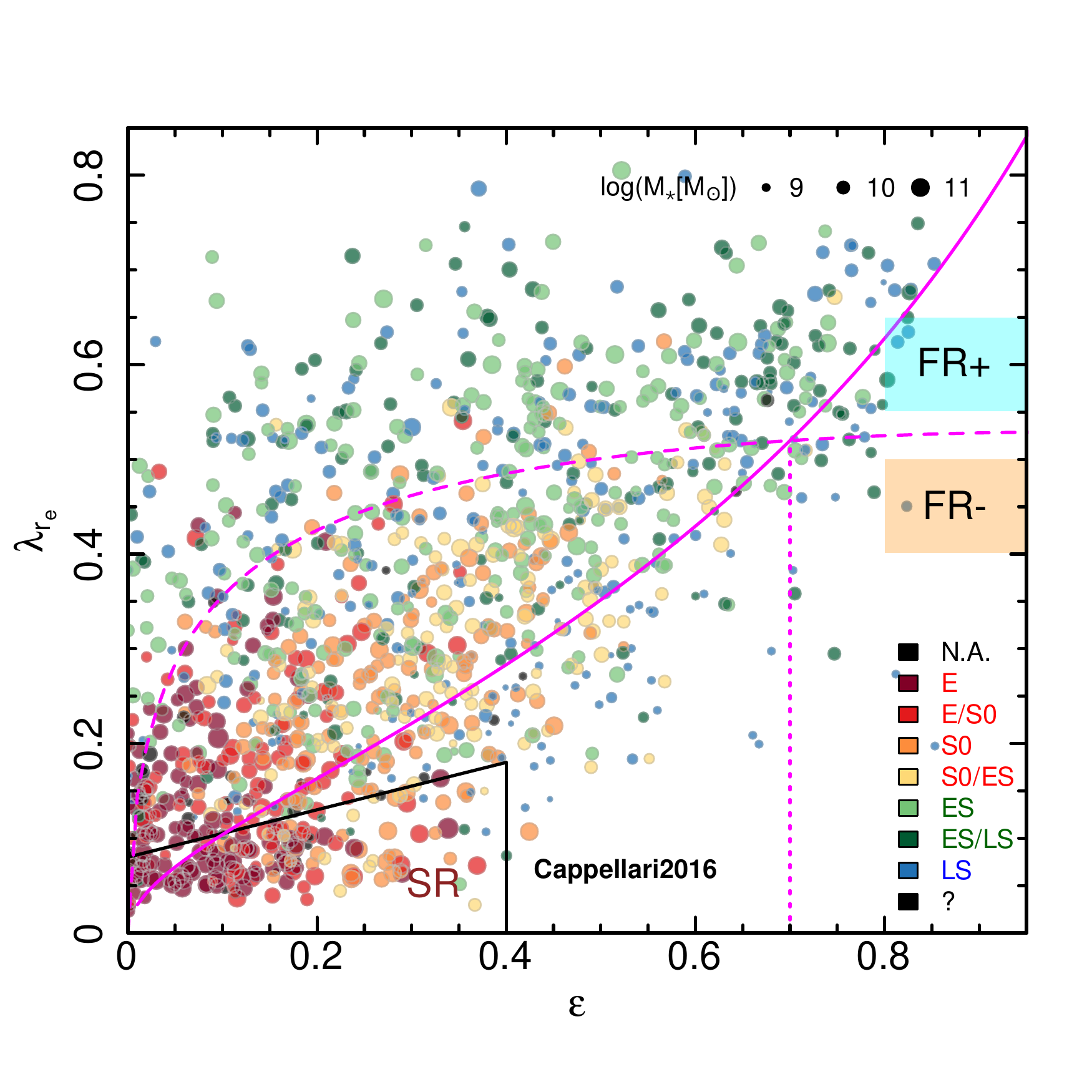}
 \caption{
 The distribution of spin parameter ($\lambda_{r_e}$) and ellipticity ($\epsilon$) 
 for the 1222 galaxies with reliable kinematic information.
 Galaxies are coded by morphological type in colour and stellar mass in size.
 The {\it black frame} on the bottom left shows the separation between fast and slow rotators defined by \citet{Cap16},
 while the {\it solid magenta line} shows the theoretical prediction for the edge-on view of 
 galaxies with constant anisotropic factor $\delta=0.7\times\epsilon_{\rm intr}$ \citep{Cap07}.
 A galaxy with $\epsilon_{\rm intr}=0.7$ viewed from different inclination angles outlines the segment between $\epsilon$ of 0 and 0.7 on the {\it magenta dashed line},
 which is selected to be the separation between FR+s and FR$-$s.
}
 \label{fig4}
\end{figure}

We show our fitting curves together with the weighted number density in Figure~\ref{fig3},
with coloured regions showing 1$\sigma$ error of the best-fit curve.
Consistent with what has been found in statistical studies based on visually-classified morphology \citep[e.g.,][]{Kel14, Mof16a},
from low to high stellar masses, the morphology of galaxies becomes more and more spheroidal, with a transition mass between disk-dominated and spheroid-dominated galaxies at around $10^{10}{\rm M_\odot}$.
It is hard to give a quantitative comparison with other studies, given the different techniques/datasets used to perform visual classification.
However, our results qualitatively agree with the SMFs of ``Spheroid Dominated'' and that of ``Disc Dominated'' 
in \citet{Kel14} and \citet{Mof16a} if LS and ES/LS labeled in Figure~\ref{fig4}
are taken as ``Disc Dominated'' galaxies.



\section{SMF divided by kinematic types}

In this section, we focus on the SMF as a function of kinematic classes
to explore the mass budget of each kinematic type to the whole galaxy population.
This can allow us to gain insights into the different physical processes responsible for (re)shaping the stellar orbits during the assembly history of galaxies.
We use the spin parameter ($\lambda_{\rm r_e}$) and ellipticity ($\epsilon$) plane to 
define different families of objects. 
We adopt the separation between FRs and SRs from \citet{Cap16} ({\it black frame} in Figure~\ref{fig4}), and we consider galaxies without available kinematic information (i.e., observed by SAMI but not fulfilling the quality cuts for stellar kinematic, ``nokin"s) as a separate type.

In order to distinguish FRs with low and high spin at fixed ellipticity, which we refer to ``FR$-$'' and ``FR$+$'', respectively,
we use the predicted relation presented by \citet[][see also Equation 14-16 in \citealt{Cap16}]{Binney05}. In particular, given that the anisotropy of a FR is bounded by a relation of the form $\delta\approx\beta_z=0.7\times\epsilon_{\rm intr}$ \citep{Cap07},
FRs with $\lambda_{\rm r_e}$ higher than predicted at a given $\epsilon_{\rm intr}$ are more isotropic than those below such a threshold
\footnote{For a detailed calculation, see Equation B1 in \citet{Ems07} and Equation 12-16 in \citet{Cap07}. We adopt $\kappa=1$ as the conversion factor between $V/\sigma$ and $\lambda_{\rm R}$. The changing of $\kappa$ value from 0.97\citep{Vds17b} to 1.1 does not affect our conclusion, simply shifting the mass fraction of 4\% from FR+ to FR$-$ if the same $\epsilon_{\rm intr}$ cut is adopted.}.
Thus, we use a value of $\epsilon_{\rm intr}=1-0.3=0.7$ 
({\it dashed magenta line} in Figure~\ref{fig4}) to divide `dynamically cold' systems (FR+) from slower axis-symmetric rotators (FR$-$). 
This is equivalent to assuming an ideal model of a galaxy with intrinsic $\epsilon$ of 0.7 and viewing it from different inclination angles. The separation line then corresponds to the different projections, according to Equation 12 in \citet[][see also \citealt{BT87}, Section 4.3]{Cap07}.
This cut is roughly consistent with the typical threshold separating visually late-type spirals from composite systems and/or bulges \citep[see also][]{Weij14,Foster17}

The plane used for our kinematic selection is shown in Figure~\ref{fig4}, with morphological types shown in different colours.
Besides a general trend from late-type dominant to early-type dominant with decreasing $\lambda_{\rm r_e}$,
it is also evident that FRs consists of all morphological types,
while SRs are mainly ellipticals or S0s with stellar mass greater than $10^{11}{\rm M_\odot}$.
This minority and mass preference of SRs is consistent with previous works focusing on fractions of galaxies with 
different kinematics \citep[e.g.,][]{Brough17,Vds17b,Graham18},
and a larger portion of the mass budget from FRs compared to SRs is also expected.

\begin{figure*}
 \includegraphics[width=\columnwidth]{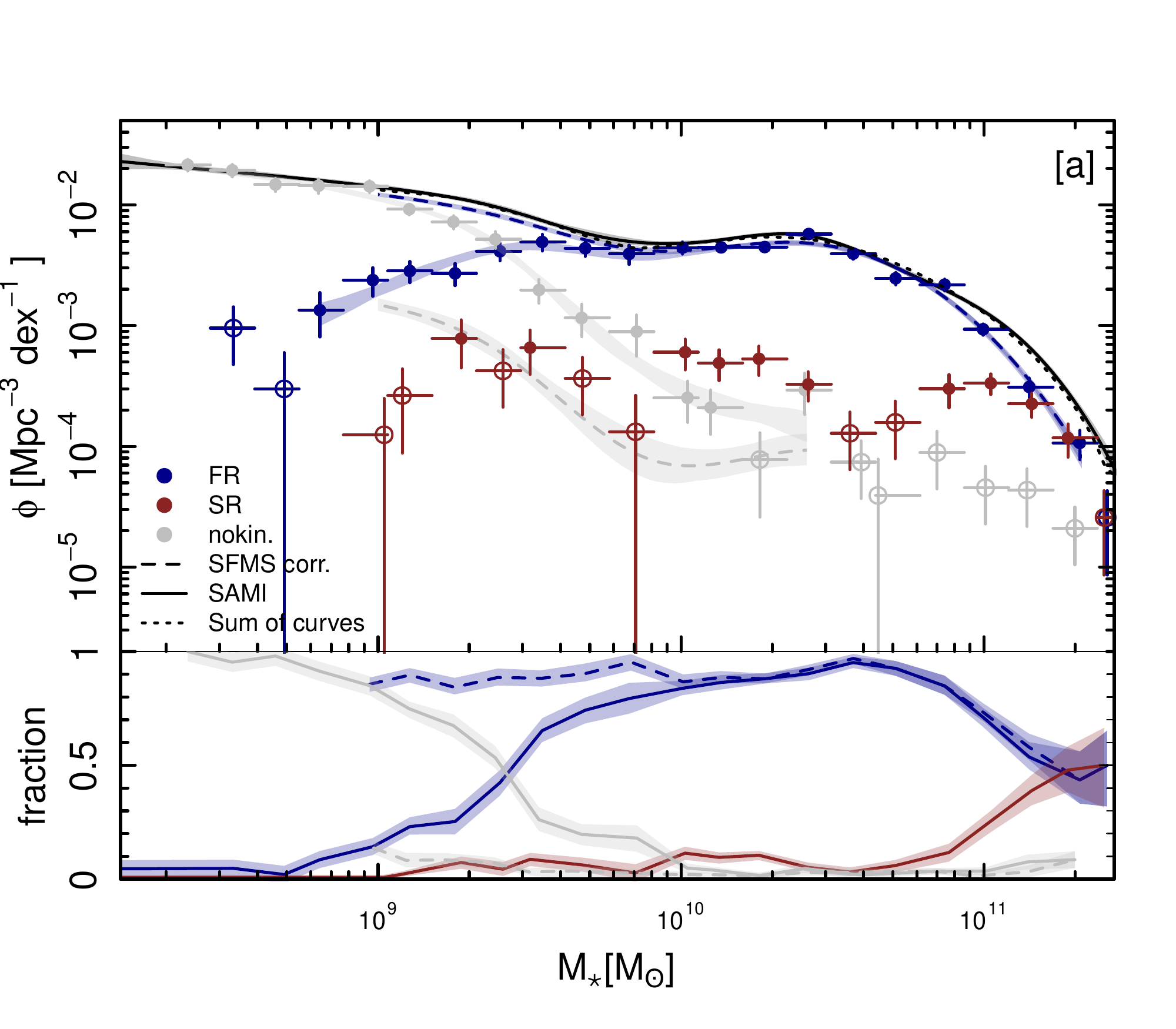}
 \includegraphics[width=\columnwidth]{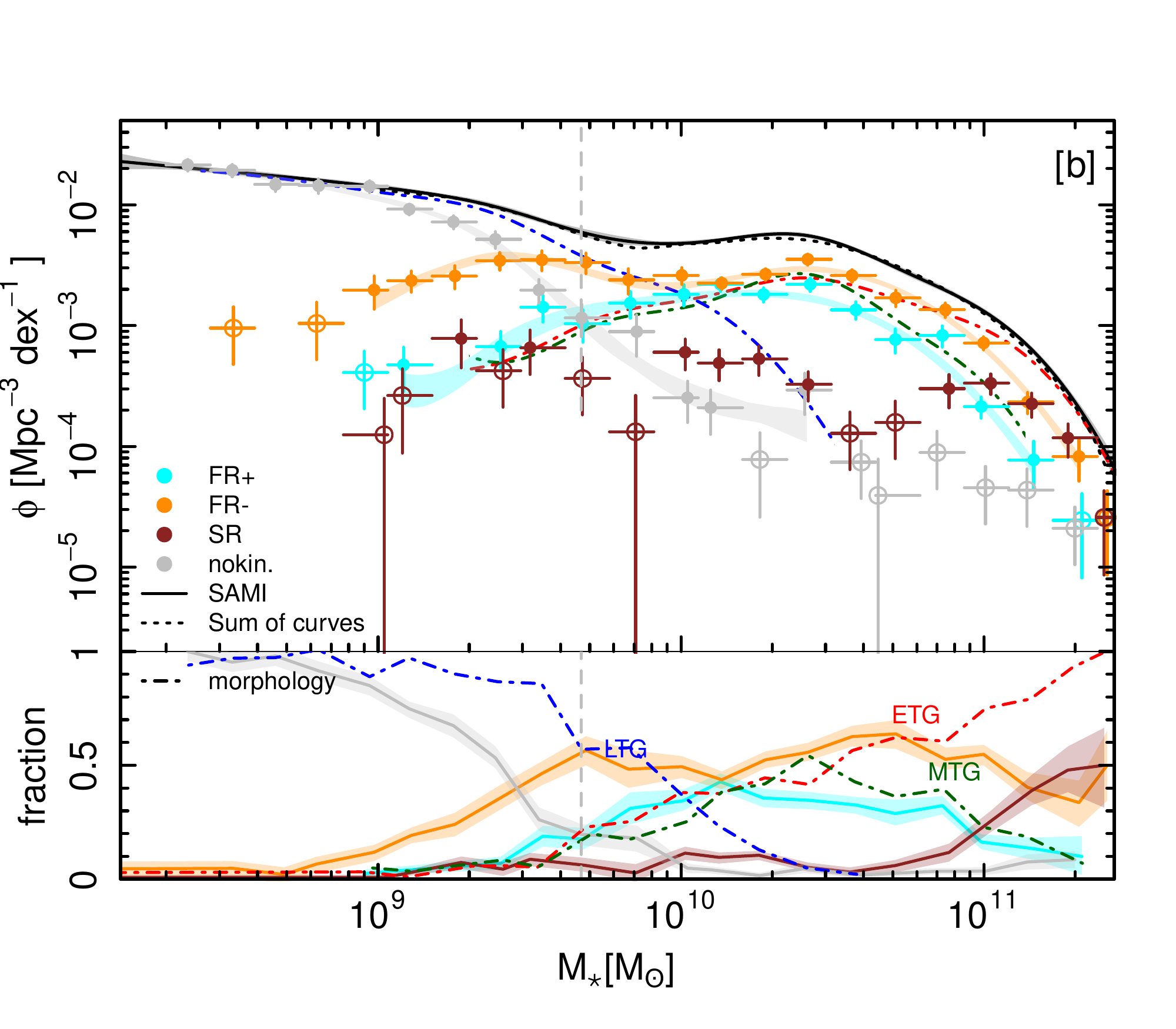}
 \caption{{\it Left:} SMFs divided by kinematic types for FRs and the number density for SRs.
 The  {\it solid black} line shows the SMF for the full sample, and  {\it gray} symbols galaxies without available kinematic information. FRs are indicated in blue and SRs in red. 
 As in Figure~\ref{fig3}, 
 points are average number density in each mass bin and 
 best-fitting curves shown only for mass bins with more than 4 galaxies
 The fraction between populations at given stellar mass (in logarithm) are shown in lower panels.
 The blue and grey dashed lines in both panels show the change introduced by assuming that all star-forming galaxies with no kinematic information are FRs. 
 {\it Right}: Same as left panel, but separating FRs+ from FRs-. The fractional distribution of morphological types is also shown for reference. In the right panel, the {\it vertical dashed line} sets a lower {\rm $M_*$} limit above which we can make an unambiguous comparison between populations.
 FRs dominate the whole galaxy sample until $10^{11}{\rm M_\odot}$ with ``warmer'' FR$-$s overwhelming ``colder'' FR+s at $M_*>10^{10}{\rm M_\odot}$, and most of massive FR+s are contributed by non-LTG galaxies.
 }
 \label{fig5}
\end{figure*}

In Figure~\ref{fig5}, we compare the incompleteness-corrected SMFs for all FRs and SRs (a), and for the three 
kinematic classes selected above (FR+, FR$-$, SR). The SMF for galaxies without kinematic information are plotted as {\it grey} lines. Due to the small number statistics of SRs, which causes the fit not make much physical sense, we show only number densities for SRs. The fraction of each population at given stellar mass is also shown in lower panels.

From Figure~\ref{fig5}a, it is clear that the number density of FRs overwhelms SRs for most stellar masses covered by SAMI, with SRs matching the fraction of FRs only at $>2\times10^{11}{\rm M_\odot}$ (see also \citealp{Kho11,Ems11,Greene17,Greene18,Vds17b, Brough17}).
We should note that this transition mass looks systematically lower than that found in \citet[][Figure 13]{Graham18} based on MaNGA galaxies. It is plausibly because we assume a unified Chabrier IMF for all galaxies throughout the study, which may lead to an underestimate of stellar mass for massive dispersion-dominated galaxies \citep{Cap12}.
Nevertheless, conversely to what is observed in the case of visual morphology between late- and early-type galaxies, there is no apparent transition point in population dominance between FR and SR classes below $10^{11.3}{\rm M_\odot}$.
This difference results from the broad overlap in kinematic properties between galaxies with different morphological type, which can already be seen from Figure~\ref{fig4}.

This difference between morphological and kinematic classification is even more dramatic if we determine the contribution of different kinematic classes to $\Omega_{*}$ by integrating the SMF for different classes for $M_*>10^{9.7}\,{\rm M_\odot}$, which roughly corresponds to the stellar mass at which our completeness in stellar kinematics drops below $\sim$80\%. We find that $\sim$82\% of the stellar mass in the local Universe is hosted in FRs, whereas only $\sim$14\% is harboured in SRs, the remaining $\sim$7\% is hosted in galaxies for which a kinematic classification was not possible. 

As shown in Figure~\ref{fig5}a, the fraction of galaxies with no kinematic information increases with decreasing stellar mass, and practically dominates our sample below $M_*\sim10^{9.5}\,{\rm M_\odot}$. This makes it unclear whether FRs dominate the whole galaxy population down to the low-mass end. To push the constraint of the ratio of FRs to SRs down to $10^9{\rm M_\odot}$,
we introduce the assumption that low-mass SFGs should be FRs. 
This is based on the expectation that star formation takes place in disks and it is supported by the fact that SRs are expected to be produced mainly by gas-poor mergers \citep[e.g.,][]{Lagos18a,Lagos18b} and 
hosted by low specific-star-formation-rate (sSFR) galaxies \citep{Naab14}.
Moreover, star-forming galaxies at this low-mass end are known to have relatively high angular momentum compared to their baryonic mass \citep{Butler17}.

Specifically, we assume galaxies with ${\rm \Delta(SFR)=SFR-SFR_{\rm MS}>-1\,dex}$ to be FRs,
where SFR is star formation rate taken from the multi-wavelength observation based MAGPHYS catalog \citep{Dav16, Driver17},
and $\rm{SFR}_{\rm MS}$ are for galaxies on the star formation main sequence (SFMS) defined by a broken linear power law approximation
\footnote{Our SFMS is $\log(SFR)=0.99\log(M_*)-9.89$ for $M_*<10^{10}{\rm M_\odot}$, 
and $\log(SFR)=0.23\log(M_*)-2.29$ for $M_*\ge10^{10}{\rm M_\odot}$.
This agrees with \citet{Guo15} and \citet{Pop19} in corresponding $\rm{M_*-z}$ range,
and the selection guarantees a fraction of $>96\%$ galaxies with kinematic information to be FRs.}. The effect of this correction to the SMFs and relative fraction of FRs in our sample is shown by the blue dashed lines in Figure~\ref{fig5}a. As expected, the fraction of FRs becomes flat down to $10^9{\rm M_\odot}$, and contribution of FRs to the stellar mass budget of the local Universe increases to 84\%.


In Figure~\ref{fig5}b, we present the SMFs and relative fractions obtained by splitting FRs into two different kinematic classes as described above. 
These two classes show a similar stellar mass distribution, with FR$-$ clearly dominating in number for $M_*>10^{10.2}{\rm M_\odot}$. It is less clear what happens for $M_*<10^{9.7}{\rm M_\odot}$: while FR$-$ seem to still be the dominant population until $10^{9.5}{\rm M_\odot}$, the number of galaxies with no kinematic information becomes significant and we cannot exclude that the decrease in the SMFs of FR+ is simply an incompleteness effect. Regardless, by integrating the SMFs of the two classes we find that FR$-$ harbour nearly a factor of $\sim$2 more mass than FR+ (i.e., 55\% vs. 27\%). 

 \begin{table}
   \caption{The percentage of the mass budget for different classes of kinematics for galaxies. Comparisons are presented for the same galaxies with seeing correction applied based on \citet{Graham18}.}
   \begin{tabular}{lccccc}
     \hline \hline
     $f_{\Omega_{*}}(M_*>10^{9.7} \rm M_\odot)$    & FR & FR+ & FR$-$ & SR & nokin \\
     \hline
     original (this work)\\({\it solid} in Figure~\ref{fig5}a) & 82\% & 27\% & 55\% & 14\% & 4\% \\
     \hline
     with seeing correction (FR)\\ of \citet{Graham18} & 82\% & 57\% & 25\% & 14\% & 4\% \\
     \hline
     with seeing correction (all)\\ of \citet{Graham18} & 88\% & 57\% & 31\% & 8\% & 4\% \\
     \hline \hline
     $f_{\Omega_{*}}(M_*>10^9 \rm M_\odot)$\\
     \hline
     original & 79\% & - & - & 14\% & 7\% \\
     \hline
     with FRs complemented\\({\it dashed} in Figure~\ref{fig5}a)  & 84\%  & -   & -   & 14\%  & 2\% \\
     \hline \hline
   \end{tabular}
   \label{tab1}
 \end{table}




It is interesting to compare visual morphology with kinematic classification. 
As shown in Figure~\ref{fig5}b, 
at stellar masses where late spirals give way to the early types ($M_*\sim10^{10}\,{\rm M_\odot}$), there is still a good fraction of dynamically cold galaxies (FR+) which are generally visually classified as two-component galaxies.
This indicates that dynamically cold galaxies defined from their kinematics are much more predominant than their morphological counterpart, i.e., the LTGs.

\begin{figure}
 \includegraphics[width=\columnwidth]{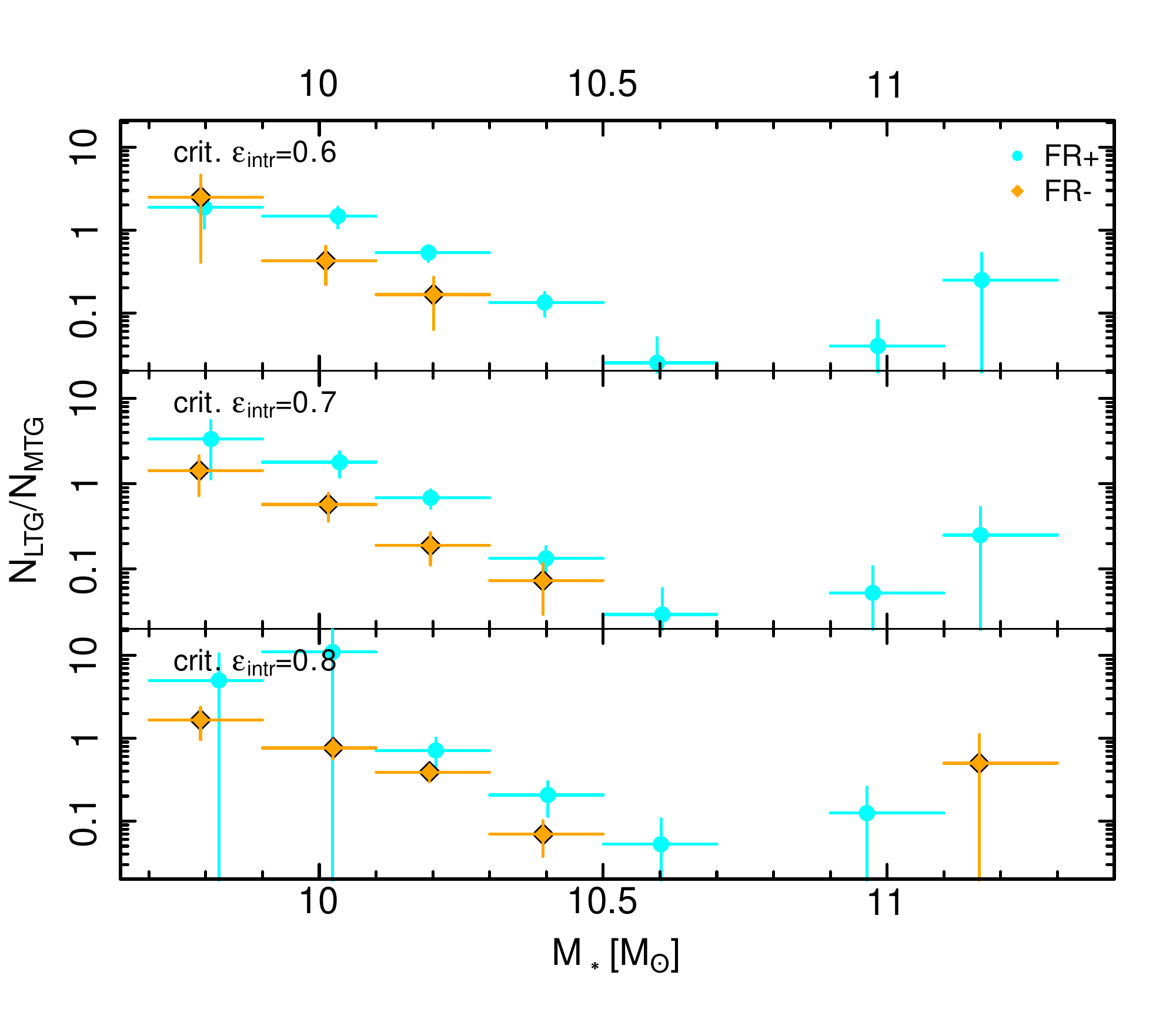}
 \caption{
 The ratio between the number of LTGs and MTGs as a function of stellar mass in groups of FR+ and FR$-$, with changing $\epsilon_{\rm intr}$ in separating dynamically-cold systems from fast rotators.
 MTGs always occupy a good fraction of massive FR+s, no matter how ``cold'' this group is defined to be.
 }
 \label{fig6}
\end{figure}

To test if this conclusion varies with the changing of our separation between FR+ and FR$-$, we check the ratio of the number of visually classified LTGs ($\rm N_{LTG}$) and MTGs ($\rm N_{MTG}$) as a function of stellar mass and the threshold used to define the two classes: i.e., 0.6, 0.7 (this work) and 0.8 (Figure~\ref{fig6}).
As  expected, FR$-$ are dominated by visually classified MTGs across nearly the entire mass range investigated here ($M_*>10^{10.2}{\rm M_\odot}$), for stellar masses greater than $\sim10^{10}{\rm M_\odot}$ the majority of FR+ would be visually classified as MTGs. While the exact value of the ratio $\rm N_{LTG}/N_{MTG}$ varies as a function of the threshold used to separate the two kinematic classes, our overall conclusions are not qualitatively affected. 

A potential issue affecting our result is the effect of beam smearing on the estimate of $\lambda_{\rm r_e}$. This could artificially reduce the value of stellar spin parameter, enhancing the importance of slow rotating systems for the stellar mass budget of the local Universe. 
To test the potential effect of beam smearing, we apply the S\'ersic-index dependent seeing correction presented by \citet{Graham18}. Although, technically, this correction is suitable only for FRs, we also check that our results do not change in case we blindly apply this correction to all our sample. The results are shown in Table~\ref{tab1}. While, as expected, the relative contribution of FR and SR to $\Omega_{*}$ is almost independent of any beam smearing effect, the situation is more complicated for the two classes of fast rotators. Once the \citet{Graham18} correction is applied, the role of FR+ and FR$-$ is reversed, with FR+ galaxies becoming the dominant population.    
While this would further reinforce our argument that dynamically cold disks are considerably more important than suggested by visual morphology, the change by almost a factor of two in the contribution to $\Omega_{*}$ of different classes of FRs highlights the challenges in interpreting individual estimates of stellar spin at face value. 




\section{Discussion and Conclusion}
In this paper, we have tested our ability to accurately reconstruct the SMFs for the SAMI Galaxy Survey, properly taking into account all the selection effects included in the original sample selection and the observations. 
We have quantified for the first time the SMFs for galaxies showing different stellar kinematic properties. 
We confirm that fast rotators (FRs) are the dominant population at almost any stellar mass. 
FRs contribute $\sim80\%$ to the total stellar mass budget, with the ratio between FRs and SRs decreasing with increasing stellar mass only for $M_*>10^{10.5}{\rm M_\odot}$.
At least one-third of the stellar mass harbored by FRs is found in systems with kinematic properties consistent with the thinnest rotating disks, 
with rotators having low stellar spin starting to dominate at $M_*>10^{10}{\rm M_\odot}$.
This is clearly in contrast with the equal contribution of spheroidal and disk components to the stellar mass budget of the local Universe as argued from visual morphology studies \citep{Mof16a, Mof16b}.

In our study,
SRs that are believed to be mostly generated from (dissipationless) mergers \citep[e.g.,][]{Jes09,Naab14,Lagos18b},
are confirmed to take up to only $<20\%$ of the mass density,
with most of them found in galaxies with $M_*>10^{10.5}{\rm M_\odot}$.
This minority suggests that the processes which completely destroy ordered rotation play a small part in the mass assembly of the average galaxy in the local Universe,
and only have effect on galaxies with $M_*>10^{10.5}{\rm M_\odot}$.

The spin of a regular rotator is intrinsically correlated to its intrinsic ellipticity \citep{Cap07}.
Therefore disk thickening and the formation of a classical bulge will cause a decrease in $\lambda_{\rm r_e}$
\citep[e.g.][and references therein]{Cap13, Cap16}, making a galaxy move downwards from the group of FR+ to FR$-$ in Figure~\ref{fig4}.
While the bulge formation could also take place at earlier times than the build-up of disk in the early assembly history of a galaxy \citep[][see also \citealt{Vds18}]{VDB98,LC16,Woo19}, nevertheless, the fact that the ratio between FR$-$ and FR+ seems to increase with mass for $M_*>10^{10}{\rm M_\odot}$ implies that the processes that efficiently thicken a disk and/or form a classical bulge are mostly efficient in this mass range.

It is also interesting to note that, for $M_*>10^{10}\,{\rm M_\odot}$, 
the number density of FR+ surpasses that of late-type spirals,
with the trend of FR+ fraction with stellar mass similar to that of MTGs (typical two-component galaxies) in morphology.
While works based on visual morphology suggest that a stellar mass of $\sim$ $10^{10}\,{\rm M_\odot}$ is a transition mass between bulge- and disk-dominant galaxies,
our results seem to imply that  many galaxies which are morphologically classified as multiple component systems including a ``bulge'' are still as dynamically cold as late-type spirals.
In other words, 
a significant fraction of the two-component galaxies are disks when using a global kinematic estimator such as $\lambda_{\rm r_e}$. 
The FR+ fraction decreases for stellar masses larger than $\sim$ $10^{10.5}\,{\rm M_\odot}$, most likely because of the increasing emergence of both disk thickening and/or the dissipationless processes creating classical bulges and/or SRs.

Lastly, we remind the reader that our classification of FR+/FR$-$, i.e., dynamically-cold/two-component disks, is based on the distribution of $\lambda_{\rm r_e}$ as a function of $\epsilon$ of galaxies, without applying any kinematic decomposition of the velocity field such as those introduced by \citet{Tabor17} and \citet{Rizzo18}. 
The next step for this type of study is to start separating different kinematic components within galaxies and to quantify their contribution to the stellar mass budget in the local Universe. 
This should provide better constraints on the different physical processes regulating the mass growth of galaxies as, at this stage, we are sensitive only to processes significantly perturbing the global stellar velocity fields. 
Given that both observational \citep{Tabor17,Tabor19,Rizzo18} and theoretical \citep{Sca10,Martig12,Wang19,Zhu18} works have clearly shown that photometric and kinematic decomposition do not always agree, we cannot blindly rely on 2D bulge-to-disk decomposition if we aim at improving our reconstruction of the accretion histories of nearby galaxies.

\section*{Acknowledgements}
We thank the anonymous referee for their constructive suggestions.
We acknowledge the helpful discussions with Dr. Angus Wright.
KG acknowledges the support from the Beijing Natural Science Foundation (Youth program) under grant no.1184015.
Parts of this research were supported by the Australian Research Council Centre of Excellence for All Sky Astrophysics in 3 Dimensions (ASTRO 3D), 
through project number CE170100013.
LC is the recipient of an Australian Research Council Future Fellowship (FT180100066) funded by the Australian Government.
DO acknowledges support by the Australia Research Council Discovery Project 160102235.
JvdS is funded under Bland-Hawthorn's ARC Laureate Fellowship (FL140100278).
SB acknowledges the funding support from the Australian Research Council through a Future Fellowship (FT140101166).
SS acknowledges the Australian Research Council Centre of Excellence for All Sky Astrophysics in 3 Dimensions (ASTRO 3D), through project number CE170100013.
JJB acknowledges support of an Australian Research Council Future Fellowship (FT180100231).
JBH is supported by an ARC Laureate Fellowship that funds Jesse van de Sande and an ARC Federation Fellowship that funded the SAMI prototype. 
MSO acknowledges the funding support from the Australian Research Council through a Future Fellowship (FT140100255). 
This research is also supported jointly by China National Postdoctoral Science Foundation (No. 2017M610696), China
Scholarship Council and the International Centre for Radio Astronomy Research.

The SAMI Galaxy Survey is based on observations made at the Anglo-Australian Telescope. 
The Sydney-AAO Multi-object Integral field spectrograph (SAMI) was developed jointly by the University of Sydney and the Australian Astronomical Observatory. 
The SAMI input catalogue is based on data taken from the Sloan Digital Sky Survey, the GAMA Survey and the VST ATLAS Survey. 
The SAMI Galaxy Survey is supported by the Australian Research Council Centre of Excellence for All Sky Astrophysics in 3 Dimensions (ASTRO 3D), 
through project number CE170100013, the Australian Research Council Centre of Excellence for All-sky Astrophysics (CAASTRO), 
through project number CE110001020, and other participating institutions. 
The SAMI Galaxy Survey website is http://sami-survey.org/ .

\label{lastpage}
\end{document}